\documentclass[12pt]{iopart}
\usepackage{iopams}
\usepackage{setstack}
\usepackage{epsf}  
\begin{document}
\title[Mean-field Monte Carlo Approach to the Dynamics of a One Pattern Model]
{Mean-field Monte Carlo Approach to the 
Dynamics of a One Pattern Model of Associative Memory}
\author{Manoranjan P. Singh\dag\ and Chandan Dasgupta\ddag\footnote{Permanent address: Department 
of Physics, Indian Institute of Science, Bangalore - 560 012, India.} \footnote[3]{To
whom correspondence should be addressed}} 
\address{\dag\ Laser Physics Division, Centre for Advanced Technology, Indore 452 013, India}
%\email{mpsingh@cat.ernet.in}
\address{\ddag\ Condensed Matter Theory Center, Department of Physics, 
University of Maryland, College Park, MD 20742-411}
\eads{\mailto{\dag\ mpsingh@cat.ernet.in}, \mailto{\ddag cdgupta@Glue.umd.edu}}
\date{10 February 2003}

\begin{abstract}
We have used a mean-field Monte Carlo method to study the zero-temperature 
synchronous dynamics of a one-pattern model 
of associative memory with random asymmetric couplings.
In the case of symmetric couplings, we find evidence for a transition 
from a spin-glass-like phase to a ferromagnet-like phase as
the acquisition strength of the stored pattern is increased from zero.
In the ferromagnetic phase, we find the existence of two types of 
phase-space structure for $m > 0$ where $m$ is the overlap of the state of the system 
with the stored pattern: a simple phase-space structure where all 
initial states with $m > 0$ flow to the attractor corresponding to the stored pattern;
and a complex phase-space structure with many attractors with their basins of attraction. 
The presence of random asymmetry  in the 
couplings results in better retrieval performance of 
the network by enhancing the size of the basin of attraction of the stored pattern
and by making the recall of memory significantly faster.
\end{abstract}
\submitto{\JPA}
\maketitle

\section{Introduction} 
Theoretical studies of neural network models of associative memory often involve the development 
of tools to study the dynamics of the network. In most simple models, 
the basic processing elements (``neurons'') are assumed to be two-state (Ising spin) variables, 
and the dynamics of the network is described by ``update rules'' that specify 
how the state of a neuron (spin) is governed by its net synaptic input (local field) 
due to the other neurons (spins) in the network. The interactions between different
neurons are specified by the synaptic matrix obtained from the learning rule employed for 
the model \cite{ch5_amit,hkp}. Dynamical studies shed light on the pattern 
recall process and its relation with the choice of the initial state, the learning rule, symmetry 
of the synaptic interaction matrix, etc. While methods of equilibrium statistical mechanics
can be profitably used \cite{ch5_amit,hkp} to analyse the behaviour of neural network models with 
symmetric synaptic connections, dynamical techniques are the {\em only} tools available
for the study of models with non-symmetric synaptic matrices. Since one is generally interested in
the behaviour of large networks, a common strategy is to move away from the 
``microscopic'' description of the dynamics of individual neurons and to derive a ``macroscopic'' 
description in terms of quantities (such as suitably defined ``order parameters") that depend 
on the states of many neurons. The crucial
question in this context is how the dynamical equations that describe the behaviour
of these macroscopic quantities are to be derived 
from the microscopic dynamics of the neurons. 

The generating functional technique 
\cite{janssen76,bausch76,martin78,dominicisPeliti78,SompolinskyZippelius,HenkelOpper91}, 
which has been used extensively to study the dynamics of spin glasses and other disordered
spin systems, provides an appropriate framework to accomplish this task. 
This technique has been used to study the synchronous dynamics of the Hopfield model 
\cite{ch5_gardner87} and its asymmetric version \cite{mpdae97, mpthesis2003}. 
Although the method allows, in principle, a calculation of all the properties of the 
network after an arbitrary number of time steps, 
it can, for all practical purposes, be used to follow the dynamics only 
for a few time steps because the the number of order parameters required in this description
increases very quickly as the number of time steps is increased. This is not satisfactory  
because, in order to analyse the retrieval properties of a neural network, one 
needs a method that allows a study of the dynamics for long times. One can, of course, use
numerical simulations for networks of finite size. However, extrapolating the results to the 
thermodynamic limit may be quite non-trivial \cite{kohring91}.

Eisffeller and Opper have developed a numerical method for studying the parallel dynamics of the 
well-known Sherrington-Kirkpatrick (SK) \cite{ch5_skmodel} model of spin glass with 
symmetric \cite{ch5_opper92} and asymmetric 
interactions \cite{ch5_opper94} in the thermodynamic limit. 
This method combines the generating functional method, which allows taking the thermodynamic 
limit exactly, and a Monte Carlo simulation of the resulting self-consistent single-spin stochastic 
dynamics. We have used this method to study the dynamics of pattern retrieval in a 
simple model of associative memory with one stored pattern. This model is essentially the same as
the SK spin-glass model with an interaction matrix that has a ferromagnetic component: the
probability distribution of each element of the interaction matrix has a positive average. 
The ferromagnetic state in this model corresponds to the stored memory, and the random part
of the interaction parameters represents the interference effects of the ``other'' memories in
Hopfield-type models with a macroscopic number of stored patterns. The relative strength of the
ferromagnetic part of the interactions plays the role of the ``acquisition strength'' of the
stored pattern. Our study leads to a characterization of the retrieval behavior of this network
as a function of this parameter. We also consider the effects of random asymmetry in the synaptic 
matrix on the retrieval performance of the model in the thermodynamic 
limit. The aim here is to shed light on the behavior of Hopfield-type models with random 
asymmetry in the synaptic connections. The main results of our study are as follows.

For symmetric couplings, we find that the finite-time dynamical behaviour of the
system exhibits a qualitative change at $J_0=1$ where $J_0$ is
the acquisition strength of the stored pattern (relative strength of the ferromagnetic part
of the interactions). This change may be described as a transition
from spin-glass-like behaviour to ferromagnet-like behavior.
In the ferromagnetic phase ($J_0>1$), we find the existence of two types of 
phase-space structure for $m > 0$ where the ``magnetization'' $m$ is the overlap of the 
state of the system with the stored pattern. For large $J_0$, the system exhibits a simple phase-space 
structure where all 
initial states with $m > 0$ flow to the attractor corresponding to the stored pattern.
The phase-space structure for smaller values of $J_0$ is complex,
with many attractors with their basins of attraction. 
In the model with symmetric couplings, the process of retrieval of memory becomes 
very slow for small values of the initial overlap. 
The presence of random asymmetry  in the 
couplings leads to an improvement in the retrieval performance of 
the network. The size of the basin of attraction of the stored pattern
increases as an antisymmetric component is introduced in the synaptic matrix. The presence
of synaptic asymmetry also decreases significantly the time the system
takes to converge to the attractor corresponding to the stored memory.

The paper is organized as follows. In Section \ref{model} we introduce the model and its basic 
properties. The generating functional technique is used to construct a mean-field theory 
for the dynamics in Section \ref{meanfield}. The results of the mean-field Monte Carlo 
simulations are presented and discussed in Section \ref{ch5_results}. The last Section 
\ref{conclusion} contains a summary of the main results and a discussion of possible connections
of these results with the behaviour of the Hopfield model with random synaptic asymmetry.      

\section{The Model}
\label{model}
The model consists of $N$ binary neurons (Ising spins) $\sigma_i=\pm 1$, where 
every neuron $\sigma_i$ is connected to all other neurons $\sigma_j$ by 
couplings $J_{ij}$:
\begin{equation}
J_{ij}=\frac{J_0}{N}\,\xi_\rmi \,\xi_j + J_{ij}^{\rm SK}\,,\quad i \ne j\,, \quad J_{ii}=0\,,
\label{Jij1}
\end{equation}
where the first term represents Hebbian learning of the binary pattern $\{\xi_i\}$ 
with $J_0$ being the  acquisition strength for the pattern \cite{hkp}. 
The second term is the coupling matrix of the SK model with random asymmetric 
interactions. As discussed in Ref. \cite{us1},  the synaptic interaction 
matrix given by Eq. (\ref{Jij1}) may be considered as a one-pattern analogue of  
the {\em tabula non rasa} scenario proposed by Toulouse, Dehaene, and Changeux \cite{ch5_tdc}. 
The couplings $J_{ij}^{\rm SK}$ are taken to be independent Gaussian random variables
for all $i < j$ with distribution 
\begin{equation}
P \left(J_{ij}^{\rm SK}\right)=\sqrt{\frac{1}{2\,\pi/N}}\, \exp\left\{
-\frac{\left(J_{ij}^{\rm SK}\right)^2}{2/N}\right\},\quad i < j\,.
\label{dist}
\end{equation}
In addition, the symmetry of the coupling matrix is given by the average symmetry parameter $\eta$:
\begin{equation}
\left[J_{ij}^{\rm SK}\,J_{ji}^{\rm SK}\right]=\eta/N \,,
\end{equation}
where the brackets denote an average over the distributions of the couplings. 
The value $\eta =1$ 
denotes  symmetric couplings whereas $\eta = -1$ corresponds to fully antisymmetric couplings. 
The case $\eta=0$ corresponds to totally uncorrelated couplings. Couplings 
with these symmetry properties can be constructed via \cite{ch5_opper94}
\begin{equation}
J_{ij}^{\rm SK}=\left[\frac{1+\eta}{2}\right]^{1/2}\,J_{ij}^{\rm s}
+\left[\frac{1-\eta}{2}\right]^{1/2}\,J_{ij}^{\rm as}\, ,
\end{equation}
where both $J_{ij}^{\rm s}$ and $J_{ij}^{\rm as}$ are independent Gaussian random variables 
for all $i < j$ with distributions same as that given by Eq. (\ref{dist}), and 
$J_{ij}^{\rm s}=J_{ji}^{\rm s}$ and $J_{ij}^{\rm as} = -J_{ji}^{\rm as}$. 
Without any loss of generality 
we can take $\xi_i = 1$ for $i=1,\ldots,N$ so that
\begin{equation}
J_{ij}=\frac{J_0}{N}+\left[\frac{1+\eta}{2}\right]^{1/2}\,J_{ij}^{\rm s}
+\left[\frac{1-\eta}{2}\right]^{1/2}\,J_{ij}^{\rm as}\, .
\label{Jij2}
\end{equation}
This form of the synaptic interaction matrix is the same as that of the asymmetric 
SK model with ferromagnetic coupling $J_0$. In this paper we consider 
the zero-temperature (noise-free) synchronous dynamics of the model:
\begin{equation}
\sigma_i(t+1)=\mathrm{sgn}\left(h_i(t)\right)\,,\quad i=1,\ldots,N,
\label{updaterule}
\end{equation}
where the local field $h_i(t)$ acting on the spin $\sigma_i$ is given by
\begin{eqnarray}
h_i(t)=\sum_{j \ne i} J_{ij}\sigma_j(t)\,,\nonumber \\
=\frac{J_0}{N}\sum_{j \ne i}\sigma_j(t)
+\left[\frac{1+\eta}{2}\right]^{1/2}\sum_{j \ne i}J_{ij}^{\rm s}\sigma_j(t)
+\left[\frac{1-\eta}{2}\right]^{1/2}\sum_{j \ne i}J_{ij}^{\rm as}\sigma_j(t)\, .
\label{localfield}
\end{eqnarray}    
In the context of the synchronous dynamics of the asymmetric Hopfield model 
considered in Refs. \cite{us1,us2}, the first term in the expression for the local field above is 
the signal term arising due to the pattern under  retrieval. The second term 
mimics the noise arising from the interference of the other stored patterns (assuming the number of
stored patterns to be a finite fraction of the number of neurons $N$). The last term 
which comes from the  antisymmetric part of the synaptic interaction matrix is the same in both 
models. At this point, it should be mentioned that in the limit of extreme dilution, the 
dynamics of the symmetrically diluted Hopfield model \cite{watkin91} can be mapped onto 
the synchronous dynamics considered here \cite{coolen2000}. Furthermore, as mentioned 
by Krauth {\it et al.} \cite{ch5_krauthNadalMezard88}, the dynamics of our model 
for $\eta=0$ is equivalent to that of the asymmetrically diluted Hopfield model which 
was introduced by Derrida {\it et al.} \cite{derrida87}. 

For $\eta \ne 1$, the long-time dynamics of the model defined above for large but finite $N$ 
is known to be rather complex \cite{cs87,somp88,gry88,bauer91,kn91,crisanti93,bastolla98}. 
In this paper, we will be concerned with the short-time dynamics of the model in the 
thermodynamic limit. The main objective here is to assess the retrieval performance 
of the network as an associative memory. To be specific, we shall study the 
initial value problem, where at time $t=0$ the spin configuration $\{\sigma_i(0)\}$ 
has a finite overlap $m_0$ with the stored pattern. Since the stored pattern in the model
is the ferromagnetic state, $\xi_i = 1$ for all $i$, the overlap $m_0$ is nothing but
the magnetization of the initial state. If 
the system evolves to a state whose overlap $m$ with the stored pattern (magnetization) 
is sufficiently close to unity, then one speaks of successful retrieval of the pattern. 
Some of the issues that are of concern in this context are:
\begin{enumerate}
\item{Retrieval quality, i.e., the closeness of the final state to the stored pattern.}
\item{Basin of attraction, i.e., the volume of phase space occupied by initial states
that converge to the attractor corresponding to the stored pattern.}
\item{Convergence time, i.e., time taken by the network to converge to the attractor
corresponding to the stored pattern.}
\end{enumerate}  
The dynamical mean-field theory described below allows us to address all these issues.
\section{Dynamical Mean-Field Theory}
\label{meanfield}
The infinite range of interactions in our model makes it amenable to exact
analysis using mean-field theory. A mean-field description involves an ``effective field", 
that depends only on some macroscopic order parameters, instead of the {\em actual} fields 
$h_i(t)$ that depend explicitly on the states of all the spins. 
However, the formulation of such a theory is highly nontrivial because of the presence of 
quenched disorder in the 
synaptic interaction matrix. The effective field for disordered models like the one considered 
here turns out to be a rather complex time-dependent random process. 
The technique of dynamic generating functional provides an 
appropriate framework for constructing the random process for the effective field. This 
random process can then be studied numerically by generating stochastic spin trajectories 
in a Monte Carlo method. 

Let us consider the statistical properties of a finite, but large number $N_T$ of 
spin trajectories of length $t_f$, at the sites $i=1,\ldots,N_T$, in a system 
where the total number $N$ of spins goes to infinity. These properties can be derived from 
the generating function $\left\langle  Z(\mathbf{l})\right\rangle_J$ for the local 
fields $h_i(t),\,i=1,\ldots,N_T$; $t=1,\ldots,t_f$.
\begin{eqnarray}
\fl \left\langle  Z(\mathbf{l})\right\rangle_J=\left\langle \Tr_{\sigma(t)}\int
\prod^N_{i=1} \prod^{t_f}_{t=1}\left\{ \rmd h_i(t)\,\Theta \left( \sigma_i(t+1)\,h_i(t)\right)
\delta\left(h_i(t)- \frac{J_0}{N}\sum_{j \ne i}\sigma_j(t) \right. \right. \right. \nonumber \\
 \left. \left. \left. \lo -\left[\frac{1+\eta}{2}\right]^{1/2} \sum_{j \ne i}J_{ij}^{\rm s}\sigma_j(t)
-\left[\frac{1-\eta}{2}\right]^{1/2}\sum_{j \ne i}J_{ij}^{\rm as}\sigma_j(t)\right)\right\} \right. \nonumber \\ 
\left. \times \exp\left( \rmi \,\sum_{t=0}^{t_f} \sum_{i=1}^{N_T} l_i(t)\,h_i(t)\right) 
\right \rangle_J.
\end{eqnarray}
Here, $\langle \cdots \rangle_J$ denotes an average over the random couplings, $\Tr_\sigma$ 
is the sum over all $2^{N\,t_f}$ possible combinations of  the spin states 
$\sigma_i(t)=\pm 1$, and $\theta(x)$ is the unit step function. By construction only 
those ``spin paths" $\sigma_i(t)$  consistent 
with the equations of motion (\ref{updaterule}) and (\ref{localfield})
contribute to $\left\langle Z(\mathbf{l})\right\rangle_J$.

The calculation of $\left\langle  Z(\mathbf{l})\right\rangle_J$ is a straightforward generalization of the 
derivation presented in Ref. \cite{ch5_opper94} for the case of $J_0=0$. 
Introducing the integral representation of the $\delta$-functions, we get
\begin{eqnarray}
\fl \left\langle  Z(\mathbf{l})\right\rangle_J\propto\left\langle \Tr_{\sigma(t)}\int
\prod_{i,\,t}\left\{ \rmd h_i(t)\rmd \hat{h}_i(t)\,\Theta \left( \sigma_i(t+1)\,h_i(t)\right)
\exp \left[\rmi \hat{h}_i(t) \left( h_i(t) \right.\right. \right. \right. \nonumber \\
\left.\left. \left. \left. \lo -\frac{J_0}{N}\sum_{j \ne i}\sigma_j(t)-
\left[\frac{1+\eta}{2}\right]^{1/2} \sum_{j \ne i}J_{ij}^{\rm s}\sigma_j(t)
-\left[\frac{1-\eta}{2}\right]^{1/2}\sum_{j 
\ne i}J_{ij}^{\rm as}\sigma_j(t) \right)\right]\right\}\right. \nonumber \\
\left.\times \exp\left( \rmi \,\sum_{i,t} l_i(t)\,h_i(t)\right) 
\right \rangle_J\,,
\end{eqnarray}
where in the last exponential only the fields $l_i(t)$ at the sites $i=1,\ldots,N_T $ 
are different from zero. Overall constants in $\left\langle Z(\mathbf{l})\right\rangle_J$ which do not depend on 
the fields $l_i(t)$ can always be recovered {\em a posteriori}, 
using the normalization relation $Z(\mathbf{l=0})=\left\langle Z(\mathbf{l=0})\right\rangle_J=1$. 
As we will find out shortly, this relation is also useful in eliminating spurious solutions. 
As noted by de Domonicis \cite{dominicis78}, since $Z(\mathbf{l=0})=1$ identically, one can 
compute directly $\langle Z \rangle_J$, the average of $Z$ over the distribution of couplings, 
thus avoiding replicas. On averaging over the disorder 
$\{J_{ij}\}$ we get
\begin{eqnarray}
\fl \left\langle  Z(\mathbf{l})\right\rangle_J \propto \Tr_{\sigma(t)}\int
\prod_{i,\,t}\left\{ \rmd h_i(t)\rmd \hat{h}_i(t)\,\Theta \left( \sigma_i(t+1)\,h_i(t)\right)\right\} \,\,
{\rme}^{\rmi \sum_{i,t}\left(l_i(t)\,h_i(t)+\hat{h}_i(t)h_i(t)\right)} \nonumber \\
\times \exp\left( -\frac{1}{2\,N}\sum_{i,j\ne i}\sum_{s,t} \left[ \hat{h}_i(t)\hat{h}_i(s)
\sigma_j(t)\sigma_j(s) + \eta \, \hat{h}_i(t)\sigma_i(s)\hat{h}_j(s)\sigma_j(t)\right]\right)\nonumber \\
\times \exp\left(-\rmi \,\frac{J_0}{N}\sum_{i,j\ne i}\sum_t\hat{h}_i(t) \sigma_i(t)\right)\,.
\end{eqnarray}
Introducing order parameters $C(t,s)$, $K(t,s)$ and $m(t)$
\begin{eqnarray}
C(t,s)& = &\frac{1}{N} \sum_j \sigma_j(t)\sigma_j(s)\,, \nonumber \\
K(t,s)& = & -\frac{\rmi}{N}\sum_j \hat{h}_j(s) \sigma_j(t)\, , \nonumber \\
m(t)& = & \frac{1}{N} \sum_j \sigma_j(t) \, ,
\end{eqnarray}
together with their conjugates $\hat{C}(t,s)$, $\hat{K}(t,s)$ and $\hat{m}(t)$, 
respectively through the identities,
\begin{equation}
\fl 1= \prod_{t,s}\left[ \int N\, \rmd C(t,s)\,\frac{\rmd \hat{C}(t,s)}{2\,\pi}\exp \left(\rmi \,N\,\hat{C}(t,s)\,
C(t,s) - \rmi \,\hat{C}(t,s)\sum_j \sigma_j(t)\sigma_j(s)\right)\right], \nonumber 
\end{equation}
\begin{equation}
\fl 1= \prod_{t,s}\left[ \int \rmi \,N\, \rmd K(t,s)\,
\frac{\rmd \hat{K}(t,s)}{2\,\pi}\exp \left(\rmi \,N\,\hat{K}(t,s)\,\rmi \,
K(t,s) - \rmi \,\hat{K}(t,s)\sum_j \hat{h}(s) \sigma_j(t)\right)\right], \nonumber 
\end{equation}
\begin{equation}
\fl 1= \prod_{t}\left[ \int N\, \rmd m(t)\,\frac{\rmd \hat{m}(t)}{2\,\pi}\exp \left( \rmi \,N\,\hat{m}(t)\,
m(t) - \rmi \,\hat{m}(t)\sum_j \sigma_j(t)\right)\right], 
\end{equation}
and neglecting terms of ${\cal O}(1/N)$, we get
\begin{eqnarray}
\fl \left\langle  Z(\mathbf{l})\right\rangle_J \propto \int \prod_t\left[N\,\rmd m(t)\,\rmd \hat{m}(t)\right]
\prod_{t,s}\left[N\,\rmd C(t,s)\,\rmd \hat{C}(t,s)\,\rmi \,N \rmd K(t,s)\,\rmd \hat{K}(t,s)\right]
\nonumber \\
\times \exp \left\{\rmi N \sum_t \hat{m}(t) m(t) + \rmi N \sum_{t,s}\left[\hat{C}(t,s)
C(t,s)+\hat{K}(t,s)\rmi K(t,s)\right]\right. \nonumber \\
\left. \lo +\sum_i \ln \left[\tilde{Z}(l_i;m,\hat{m},C,\hat{C},K,\hat{K})\right]\right\}\, ,
\label{saddlez}
\end{eqnarray}
where the single-site partition function $\tilde{Z}_i$ is given by is  
\begin{eqnarray}
\fl \tilde{Z}(l_i;m,\hat{m},C,\hat{C},K,\hat{K}) \propto \Tr_{\sigma_i(t)}\int
\prod_t\left\{ \rmd h_i(t)\rmd \hat{h}_i(t)\,\Theta \left( \sigma_i(t+1)\,h_i(t)\right)\right\}\nonumber \\
\times \exp\left\{\rmi \sum_t\left(l_i(t)\,h_i(t)+\hat{h}_i(t)h_i(t)\right)\right\} \nonumber \\
\times \exp\left\{ -\rmi J_0 \sum_t m(t) \hat{h}_i(t) -\rmi \sum_t \hat{m}(t)\sigma_i(t) \right. \nonumber 
\\ \left. \lo -\sum_{s,t}\left( \frac{1}{2}C(t,s) \hat{h}_i(t)\hat{h}_i(s)
+\rmi \hat{C}(t,s)\sigma_i(t)\sigma_i(s)\right.\right. \nonumber \\ 
\left. \left. \lo +\frac{\rmi\eta}{2}K(t,s)\hat{h}_i(t)\sigma_i(s) 
+\rmi \hat{K}(t,s)\hat{h}_i(s)\sigma_i(t) \right) \right\}.
\end{eqnarray}

At this stage the dynamical variables are decoupled with respect to their site 
index $i$. The exponent in Eq. (\ref{saddlez}) is of the form $N\,F(m,\hat{m},C,\hat{C},K,\hat{K})$. 
Therefore, in the limit $N\rightarrow \infty$ the integration over $m(t),\,\hat{m}(t),\,
C(t,s),\,\hat{C}(t,s),\,K(t,s),\, {\rm and}\, \hat{K}(t,s)$ can be performed using the saddle-point 
method. The stationary values of the order parameters are found from the following set 
of equations:
\begin{equation}
\hat{m}(t)=\frac{J_0}{N} 
\sum_i \left\langle \hat{h}_i(t) \right\rangle_{\tilde{Z}_i}\,, \label{saddlemhat}
\end{equation}
\begin{equation}
m(t)=\frac{1}{N} 
\sum_i \left\langle \sigma_i(t) \right\rangle_{\tilde{Z}_i} \,, \label{saddlem}
\end{equation}
\begin{equation}
\hat{C}(t,s)=-\frac{\rmi}{2\,N} 
\sum_i \left\langle \hat{h}_i(t)\hat{h}_i(s) \right\rangle_{\tilde{Z}_i}\,, \label{saddlechat}
\end{equation}
\begin{equation}
C(t,s)=\frac{1}{N} 
\sum_i \left\langle \sigma_i(t)\sigma_i(s) \right\rangle_{\tilde{Z}_i}\,, \label{saddlec}
\end{equation}
\begin{equation}
\hat{K}(t,s)=-\frac{\rmi \,\eta}{2\,N} \sum_i \left\langle 
\hat{h}_i(t)\sigma_i(s) \right\rangle_{\tilde{Z}_i}\,, \label{saddlekhat}
\end{equation}
\begin{equation} 
K(t,s)=-\frac{\rmi}{N} 
\sum_i \left\langle \hat{h}_i(s) \sigma_i(t) \right\rangle_{\tilde{Z}_i}\,. \label{saddlek}
\end{equation} 
%\end{eqnarray}
In these equations, $\langle \cdots \rangle_{\tilde{Z}_i}$ denotes an average with respect to the single-site 
partition function with $l_i(t)=0$. Eqs. (\ref{saddlemhat}) and (\ref{saddlechat}) have only the 
trivial solutions, 
\begin{eqnarray}
\hat{m}(t)&=&0\,,\\
\hat{C}(t,s)&=&0\,,
\end{eqnarray}
as any other solution would violate the normalization $Z(\mathbf{l}=0)=1$. 

From Eqs. (\ref{saddlekhat}) and (\ref{saddlek}) we get
\begin{equation}
\hat{K}(t,s)=\frac{\eta}{2} K(s,t)\,.
\end{equation}
As we will see below, $K(t,s)$ is the average response of the magnetization at time 
$t$ with respect to a small variation of the external field at time $s$. We are interested 
in the solutions that respect causality, i.e.,
\begin{equation}
K(t,s)=0\quad {\rm for} \quad s \ge t \,.
\label{causalk}
\end{equation}

Once again using the normalization property of $Z(\mathbf{l})$, we omit the 
single-site partition functions with $l_i=0$ to get
\begin{eqnarray}
\left\langle  Z(\mathbf{l})\right\rangle_J&\propto& \prod_{i=1}^{N_T}\Tr_{\sigma_i(t)}\int
\prod_t\left\{ \rmd h_i(t)\rmd \hat{h}_i(t)\,\Theta \left( \sigma_i(t+1)\,h_i(t)\right)\right\} \nonumber \\
&&\exp\left\{\rmi \sum_t \left( l_i(t)\,h_i(t)+
\hat{h}_i(t)h_i(t)-\rmi \hat{h}_i(t)J_0 m(t)\right)  \right. \nonumber \\
&&  \left.  -\frac{1}{2}\sum_{s,t} C(t,s) \hat{h}_i(t)\hat{h}_i(s)
-\rmi \eta \sum_{s,t}K(t,s)\hat{h}_i(t)\sigma_i(s) \right\}\,.
\label{effectivez}
\end{eqnarray}

The generating functional (\ref{effectivez}) describes a system of $N_T$ {\em noninteracting} 
spins. It can be rewritten in a form where each spin is 
coupled to an effective field. In order to accomplish this we linearize the 
quadratic terms in $\hat{h}_i(t)$ by introducing Gaussian random variables $\phi_i(t)$, 
with zero mean and covariance 
$\left\langle \phi_i(t)\phi_i(s)\right\rangle_\phi = C(t,s)$, independently for each site $i$. Using the identity
\begin{eqnarray}
\exp\left\{-\frac{1}{2} \sum_{s,t}\left\langle \phi_i(t)\phi_i(s)\right\rangle_\phi 
\hat{h}_i(t)\hat{h}_i(s)\right\} = \left\langle\exp\left\{-\rmi \sum_t \phi_i(t)
\hat{h}_i(t) \right\}\right\rangle_\phi
\end{eqnarray}
where $\langle \cdots \rangle_\phi$ denotes an average over the time dependent Gaussian 
random variables $\phi_i(t)$, we get
\begin{eqnarray}
&\left\langle  Z(\mathbf{l})\right\rangle_J&\propto \prod_{i=1}^{N_T}\left\langle\Tr_{\sigma_i(t)}\int
\prod_t\left\{ \rmd h_i(t)\rmd \hat{h}_i(t)\,\Theta 
\left( \sigma_i(t+1)\,h_i(t)\right)\right\} \right. \nonumber \\
&&\left. \exp\left\{\rmi \sum_t \left( l_i(t)\,h_i(t)+ \hat{h}_i(t)h_i(t)-
\rmi \hat{h}_i(t)J_0 m(t)\right) \right. \right. \nonumber \\
&& \left. \left.  -\rmi\sum_{t}\phi_i(t) \hat{h}_i(t)
-\rmi\eta \sum_{s,t}K(t,s)\hat{h}_i(t)\sigma_i(s) \right\}\right\rangle_\phi\,.
\label{effectivezphi}
\end{eqnarray}
On integrating over the auxiliary fields $\hat{h}_i(t)$, we get to the result 
\begin{eqnarray}
\fl \left\langle  Z(\mathbf{l})\right\rangle_J \propto & \prod_{i=1}^{N_T}\left\langle\Tr_{\sigma_i(t)}\int
\prod_t\left\{ \rmd h_i(t)\,\Theta \left( \sigma_i(t+1)\,h_i(t)\right)\right\} 
\exp\left\{\rmi \sum_t l_i(t)\,h_i(t) \right\} \right.\nonumber \\
& \left. \prod_t\delta\left(h_i(t)-J_0 m(t)-\phi_i(t) -\eta \sum_s K(t,s)\sigma_i(s) \right)\right\rangle_\phi\,.
\label{zfinal}
\end{eqnarray}

This representation of the generating function implies that the dynamics of the spin 
system given by Eq. (\ref{updaterule}) is described by the uncorrelated system of 
stochastic dynamical equations:
\begin{equation}
\sigma_i(t+1)=\mathrm{sgn}\left(h_i(t)\right)\,,
\label{effectivedynamics}
\end{equation}
with
\begin{equation}
h_i(t)=J_0 m(t) + \phi_i(t) + \eta\sum_{s < t} K(t,s) \sigma_i(s)\,.
\label{effectiveh}
\end{equation}
The first term in the ``effective" local field is a simple disorder-free mean field 
term, the second term is a non-white Gaussian noise, while the third term represents 
a retarded self-interaction.

The order parameters given by Eqs. (\ref{saddlem}), (\ref{saddlec}), and (\ref{saddlek}) 
can be rewritten in terms of the Gaussian averages:
\begin{eqnarray}
m(t)&=&\left\langle \sigma(t) \right\rangle_\phi\,, \label{mphi}\\
C(t,s)&=&\left\langle \phi(t)\phi(s) \right\rangle_\phi 
=\left\langle \sigma(t)\sigma(s) \right\rangle_\phi\,, \label{cphi}\\
K(t,s)&=&-\rmi \left\langle \hat{h}(s) \sigma(t) \right\rangle_\phi
=\left\langle \frac{\partial}{\partial{\phi(s)}}\sigma(t) \right\rangle_\phi\,. \label{kphi}
\end{eqnarray}
Eq. (\ref{kphi}) clearly brings out the physical interpretation of $K(t,s)$ as a response function. 
However, it would be highly inconvenient to use this relation to evaluate $K(t,s)$ as it requires 
a calculation of the average of the partial derivative. Therefore, we use a discrete version of  
Novikov's theorem \cite{ch5_opper94, hanggi78} to express this quantity in terms of the 
correlation function $\langle \sigma(t) \phi(s) \rangle$ which is easier to estimate:
\begin{equation}
\langle \sigma(t) \phi(s) \rangle = \sum_{\tau=0}^t K(t,\tau) C(\tau,s)\,.
\label{novikov theorem}
\end{equation}
We note that Eq. (\ref{novikov theorem}) holds independently of the value of the asymmetry parameter 
$\eta$. On the other hand, a fluctuation dissipation theorem, which would enable us to 
express $K(t,s)$ directly in terms of $C(t,s)$, is not available for the asymmetric 
synaptic interaction matrix considered here.

\section{Results of Monte Carlo Simulation}
\label{ch5_results}
The single spin equations (\ref{effectivedynamics}) and (\ref{effectiveh}) can 
be used to calculate exact averages for $N \rightarrow \infty$ by expressing the spin 
variables as an explicit function of the Gaussian fields $\phi(t)$ and performing 
integrations weighted by the multivariate Gaussian measure. This 
integration is most conveniently performed by a Monte Carlo process, where a 
sequence of Gaussian random numbers with respect to  the covariance $C(t,s)$ is 
generated and a trajectory of spins $\sigma(t)$ is created via Eqs. (\ref{effectivedynamics}) 
and (\ref{effectiveh}). The necessary average at each 
time step is estimated by summing over a large number $N_T$ of trajectories. $N_T$ should 
not be confused with $N$, the number of spins in the model, which 
tends to infinity. We closely follow the algorithm for the Monte Carlo simulation 
presented in Ref. \cite{ch5_opper94}. We take $\sigma^k(0)=\pm 1$ with probabilities 
$(1 \pm m_0)/2$, respectively, for all $k=1,\ldots,N_T$, where $m_0$ is the initial 
overlap with the stored pattern. In all our simulations we have taken $N_T=10^6$. 
Although, in most cases we have restricted the temporal range to $t_f < 200$, occasionally 
we have gone beyond this range to bring out some qualitative  features of the dynamics. 

\subsection{Symmetric Couplings: $\eta=1$}
\begin{figure}[htb]
\begin{center}
\epsfxsize=10cm
\epsfysize=8cm
\epsfbox{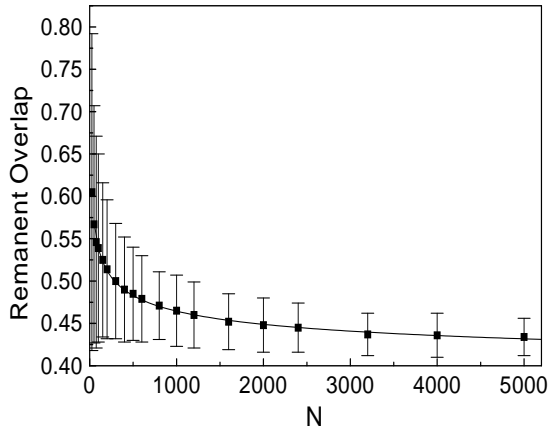}
\end{center}
\caption{\label{fig1} Remanent overlap for the even time dynamics obtained by numerical 
simulation for $J_0=0.8$, $\eta=1$, $m_0=1$ and various values of the network size $N$. 
The full curve denotes the best fit of the results to the form of Eq. (\ref{fitn}).}
\end{figure}

\begin{figure}[htb]
\begin{center}
\epsfxsize=0.49\textwidth
\epsfysize=8cm
\epsfbox{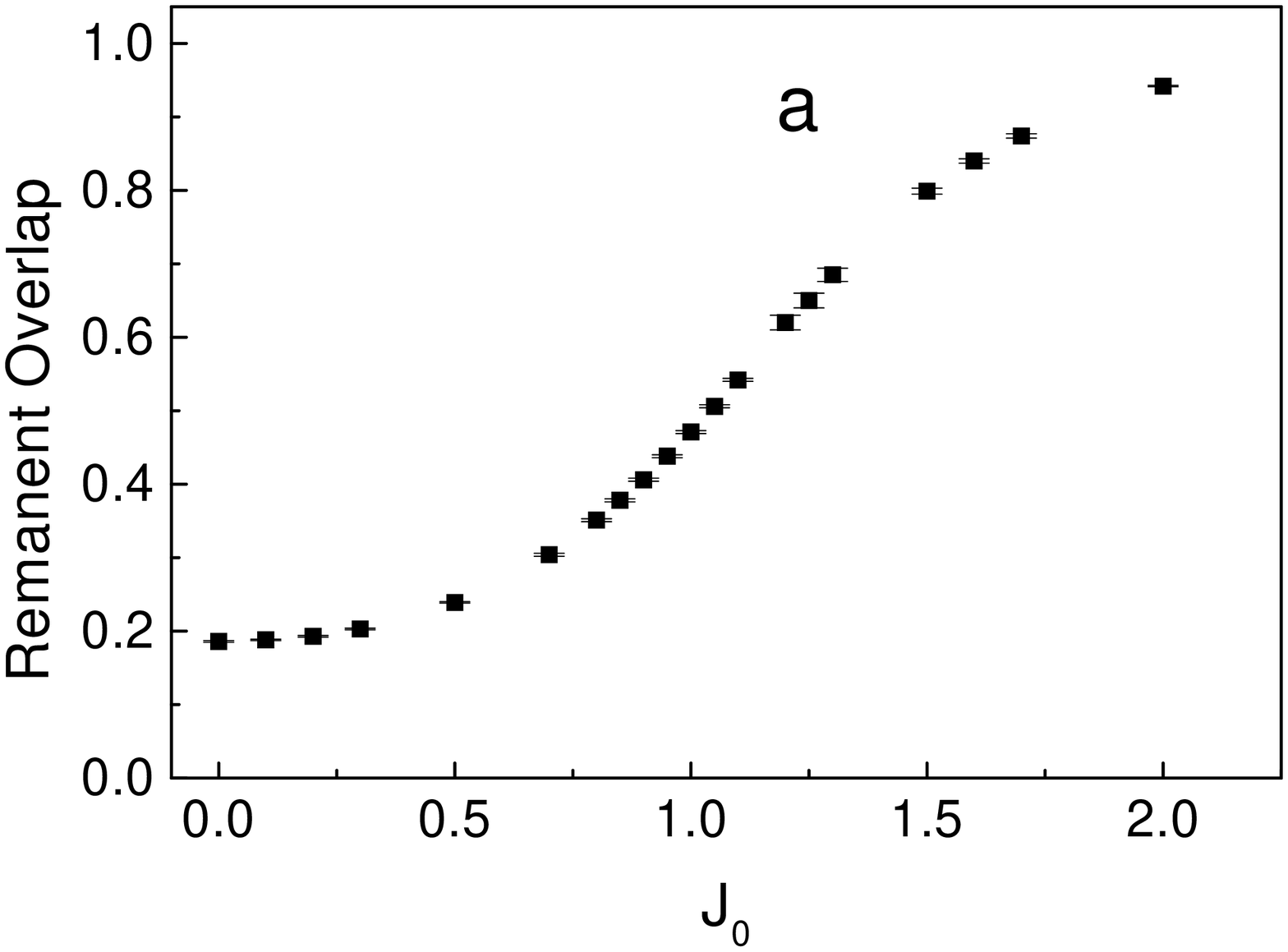}
\epsfxsize=0.49\textwidth
\epsfysize=8cm
\epsfbox{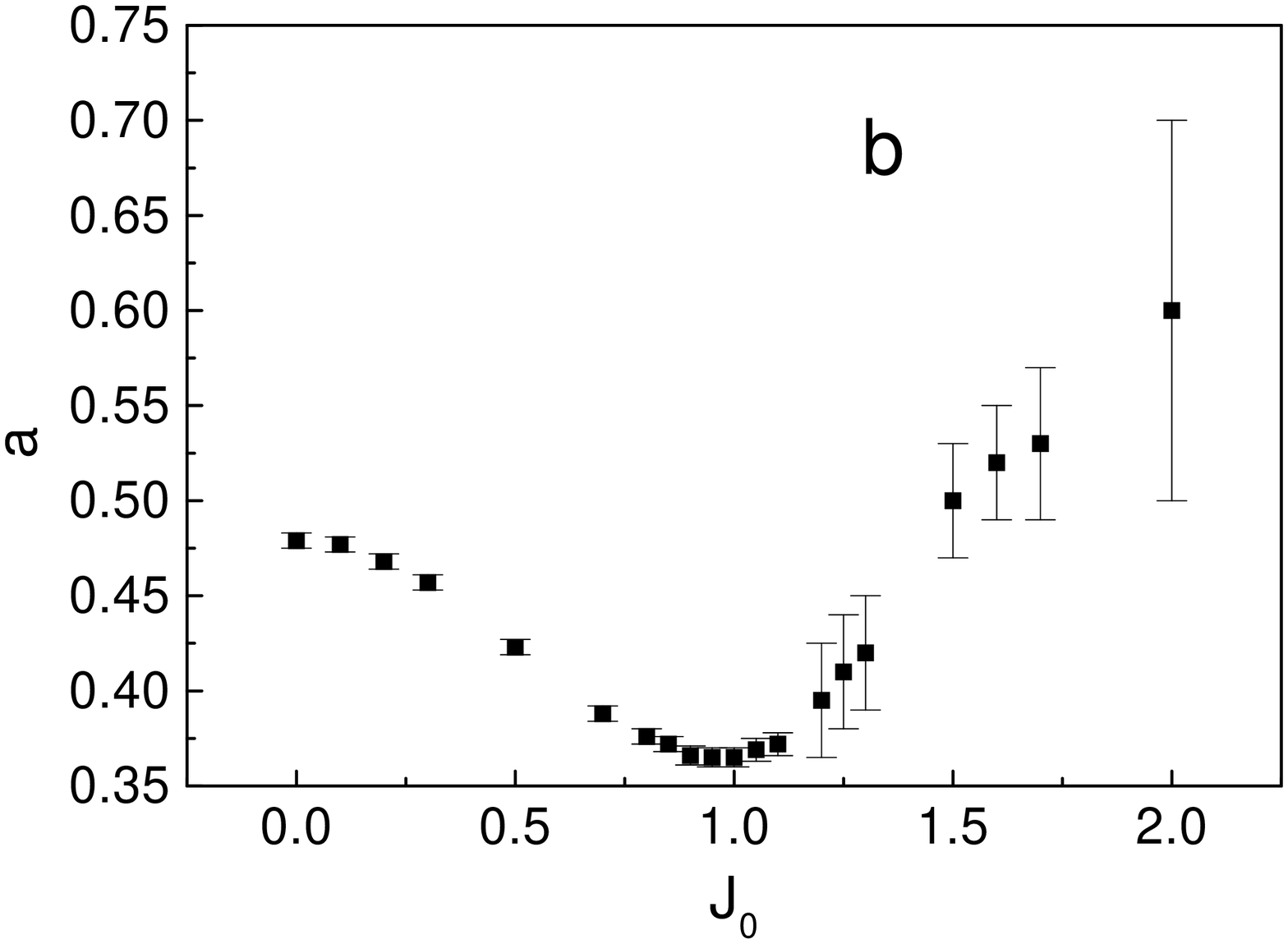}
\end{center}
\caption{\label{fig2} Results of fitting the mean-field Monte Carlo data for the evolution 
of the overlap, $m(t)$, at even times to Eq. (\ref{fittingfunction1}) for $\eta=1$ and $m_0=1$. 
(a) Remanent overlap $m_{\infty}$ as a function of the acquisition strength $J_0$. 
(b) The exponent $a$ of the power law as a function of $J_0$.}
\end{figure}

We first look at the stability of the stored pattern $\{1,1,\ldots,1\}$. Accordingly, 
we start from $\sigma^k(0)=1$ for $k=1,\ldots,N_T$, i.e. $m_0=1$, and evaluate $m(t)$ 
for 100 time steps. Since the system would settle down, in general, to a limit cycle of 
length 2, we analyze the dynamics for even and odd times 
separately. We find that the results of simulations in both the cases can be fitted well 
by the function
\begin{equation}
m(t)=m_{\infty}+\mathrm{const} \times t^{-a}\,,
\label{fittingfunction1}
\end{equation} 
where the parameters $a$ and $m_{\infty}$ are functions of $J_0$, the acquisition strength 
of the stored pattern. $m_{\infty}$ is the extrapolated value of the overlap for 
$t_{\rm even}$ (or $t_{\rm odd}) \rightarrow \infty$. For estimating uncertainties in the 
values of  the fitting parameters, $m_{\infty}$, const, and $a$, we take the uncertainty in
the values of $m(t)$ to be $\Delta m(t) = 10^{-3}$, i.e., $\sim O(1/\sqrt{N_T})$ at all 
time steps \cite{ch5_opper94}. In order to have a cross-check on the results obtained from 
the mean-field Monte Carlo procedure, 
we did numerical simulation of Eqs. (\ref{updaterule}) and (\ref{localfield}) for $J_0=0.8$ 
and $m_0=1$. The calculations were done on finite samples of $N$ sites $(25 \le N \le 5000)$ 
and the overlap $m(t)$ was averaged over 100 to 2$\times 10^5$ samples. 
In Fig. \ref{fig1}, we plot $m_{\infty}(N)$, the remanent overlap for the even time 
dynamics. By fitting $m_{\infty}(N)$ to the function
\begin{equation}
m_{\infty}(N)  = m_{\infty}+ \mathrm{const}\times N^{-b}\,,
\label{fitn}
\end{equation}
we get $m_{\infty}=0.36 \pm 0.05$ which is in good agreement with $m_{\infty}=0.36 \pm 0.02$ 
obtained by the mean-field Monte Carlo method described above.
We show in Fig.~\ref{fig2} the remanent overlap $m_{\infty}$ and the exponent $a$ 
of the power law decay of the even time dynamics in Eq. (\ref{fittingfunction1}) as functions 
of $J_0$. The plots show clear evidence for a ``transition" near $J_0=1$: the rate of increase
of the remanent overlap $m_\infty$ with increasing $J_0$ is maximum near $J_0=1$, and the exponent
$a$ has a minimum at the same value of $J_0$. This ``transition'' is from spin-glass-like
to ferromagnet-like behaviour. The ``spin-glass'' phase for $J_0<1$ has a small 
value of the remanent overlap, arising due to the non-ergodic relaxation of the system 
through a complex energy landscape, which prevents it from reaching the equilibrium state 
corresponding to $m=0$. The system gets frozen to a cycle of length two which can be 
characterized by the remanent overlaps $m_{\infty}^{\rm even}$ and $m_{\infty}^{\rm odd}$ 
of the even and odd time dynamics, respectively. (Wherever we discuss the even time dynamics 
alone, we drop the superscript. Thus, $m_{\infty}^{\rm even}$ and $m_{\infty}$ both refer to the 
remanent overlap for the even time dynamics). For $J_0=0$, $m_{\infty}^{\rm even} = 0.186 \pm 0.001$ 
whereas $m_{\infty}^{\rm odd}=0$ (precisely, $O(10^{-3})$ which is the inherent level of errors 
involved in the calculation). Both of the remanent overlaps increase with $J_0$. 
Coming back to the even time dynamics we find in Fig. 2(b) that the relaxation becomes 
slower as $J_0$ is increased in the spin-glass phase. On the other hand, in the ``ferromagnetic'' 
phase ($J_0>1$), the system relaxes faster for higher value of $J_0$. The behavior is consistent with 
the physical intuition that  $J_0$ is the strength of the  ferromagnetic coupling 
which opposes the decay of the system to a small value of $m_{\infty}$ in the 
spin glass phase, while helping the system to have a large $m_{\infty}$ in the ferromagnetic phase.
We should emphasize that the ``transition'' mentioned above reflects a qualitative change in
the short-time dynamical behavior of the system -- it does not necessarily correspond to a 
phase transition in the thermodynamic sense. It is, however, interesting that the value of 
$J_0$ where this change in the dynamics occurs is consistent with the phase diagram 
of the SK model with ferromagnetic interactions \cite{fischerhertz91}, obtained  
by the replica theory, which shows a transition from the spin-glass phase to a 
ferromagnetic phase (with replica symmetry broken) at $J_0=1$.

Strictly speaking, the stored pattern is not absolutely stable for any finite value of $J_0$: the
remanent overlap $m_\infty$ is always less than unity. However, 
this does not prevent the network from performing as an associative memory. For  sufficiently 
large values of $J_0$, we can have $m_{\infty}$ very close to unity (e.g., for $J_0=2.0$, 
$m_{\infty}=0.942 \pm 0.001$). $m_{\infty}$ can be made as close to unity as we wish by 
increasing $J_0$. A similar scenario exists in the Hopfield model, where for extensive 
loading of memory, the stored patterns are not fixed points of the dynamics. In the Hopfield 
model, too, we have retrieval fixed points which can be made closer to the respective 
stored patterns by reducing the memory loading level of the network. What matters 
in both the models is the ratio of the strengths of the signal and the noise terms in 
the expressions for the local fields. This substantiates the analogy of $J_{ij}^{\rm s}$ 
with the noise arising from stored patterns other than the one 
under retrieval in the Hopfield model.

\begin{figure}[htb]
\begin{center}
\epsfxsize=10cm
\epsfysize=8cm
\epsfbox{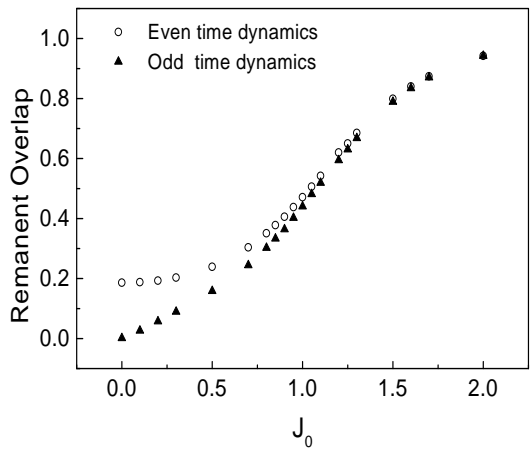}
\end{center}
\caption{\label{fig3} Remanent overlaps for the even-time and odd-time dynamics as functions 
of the acquisition strength $J_0$ for $\eta=1$ and $m_0=1$.}
\end{figure}

As mentioned earlier, the synchronous dynamics, in general, takes the system to a 
limit cycle of length 2. For the network to function as an associative memory, it is 
desirable that the values of the overlap to which the network settles down at even and 
odd times are not very different. In Fig. \ref{fig3} we plot together $m_{\infty}^{\rm even}$ and  
$m_{\infty}^{\rm odd}$ as functions of $J_0$. It is evident that for the values of the 
acquisition strength that are of interest ($J_0 > 1$), the difference between the two remanent 
overlaps are very nominal. We, therefore, concentrate only on the even time dynamics hence after.

\begin{figure}[htb]
\begin{center}
\epsfxsize=10cm
\epsfysize=8cm
\epsfbox{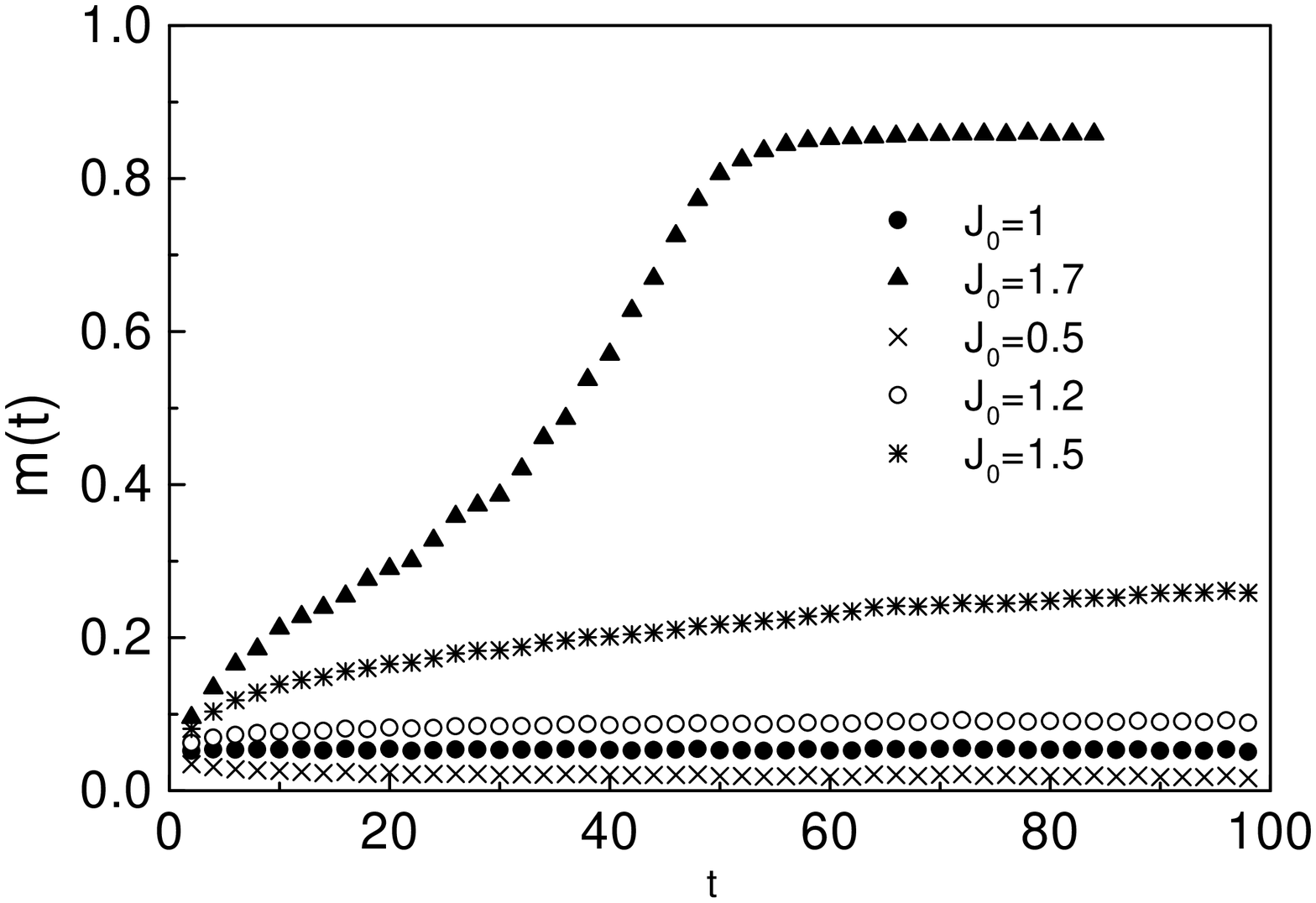}
\end{center}
\caption{\label{fig4} Evolution of the overlap $m(t)$ at even times for different values 
of $J_0$ for $\eta=1$ and $m_0=0.05$.}
\end{figure}

\begin{figure}[htb]
\begin{center}
\epsfxsize=10cm
\epsfysize=8cm
\epsfbox{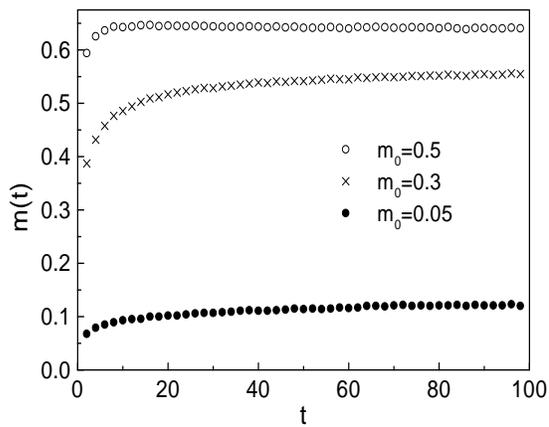}
\end{center}
\caption{\label{fig5} Evolution of the overlap $m(t)$ at even times for different values 
of $m_0$, the initial overlap with the stored pattern. The plots are for $\eta=1$ and $J_0=1.3$.}
\end{figure}

\begin{figure}[htb]
\begin{center}
\epsfxsize=10cm
\epsfysize=8cm
\epsfbox{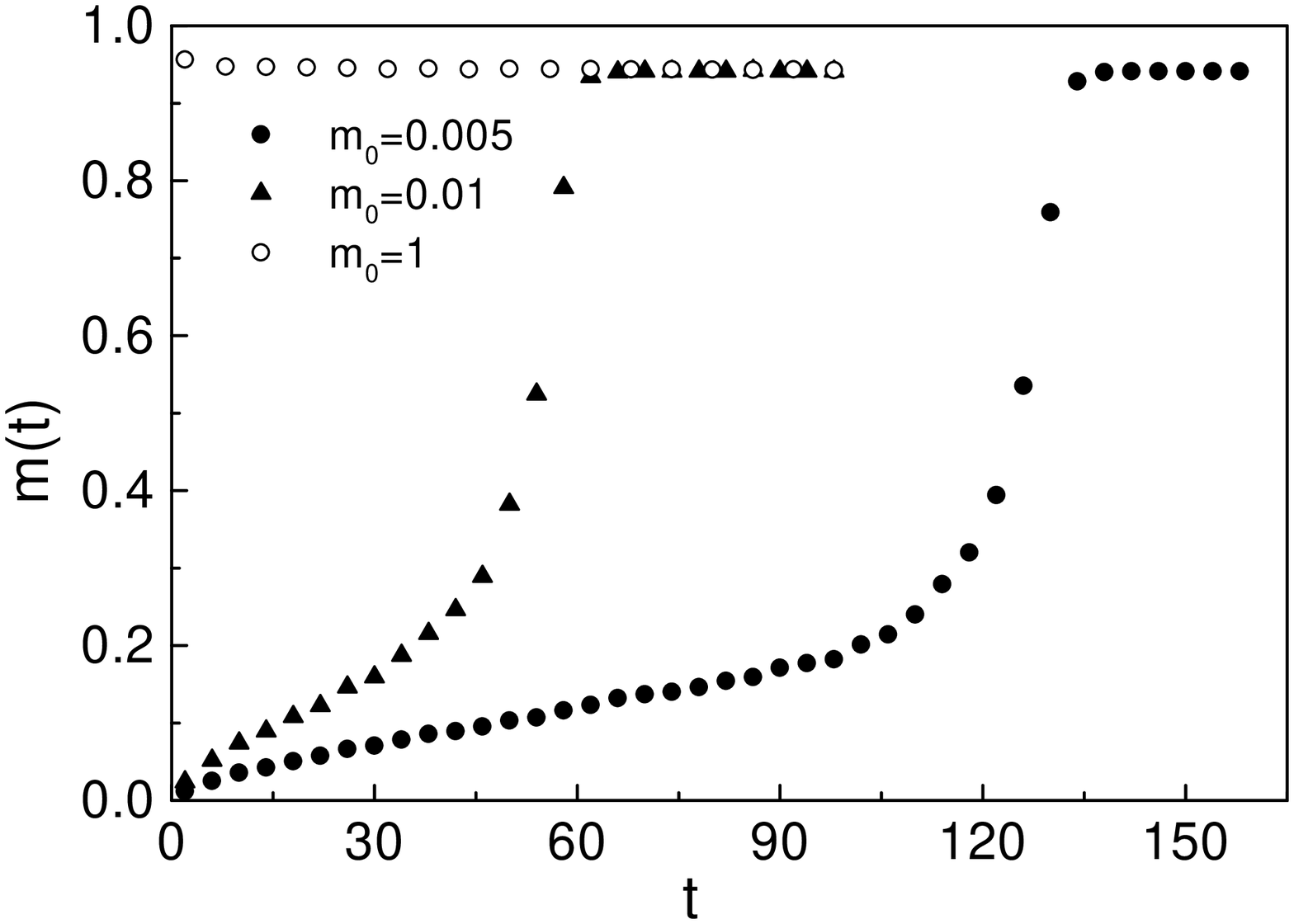}
\end{center}
\caption{\label{fig6} Same as Fig. \ref{fig5} for $J_0=2$.}
\end{figure}

For the network to function as an associative memory, it also desirable that the attractor 
(limit cycle) corresponding to the stored pattern has a large basin of attraction, 
i.e., a large number of 
initial configurations having a finite overlap with the stored pattern (i.e., with $m_0 \ne 0$) 
should converge to this attractor. We, therefore, have 
studied the dynamics of the system with $m_0=0.05$, which is well below the remanent overlap 
for all values of $J_0$. We find that $m(t)$ increases with time  if 
$J_0 > 1$, decreases with time if $J_0<1$, and remains nearly constant for $J_0=1$. 
This behavior, shown in Fig. \ref{fig4}, also suggests a transition at $J_0=1$.  
Even for the values of $J_0 >1$, two different kinds of behavior of $m(t)$ are possible. 
For relatively smaller values of $J_0$, the 
value of $m_{\infty}$ depends on the initial overlap $m_0$, e.g., for $J_0=1.3$ the values of 
$m_{\infty}$ for $m_0=0.05,\,0.3,{\rm and}\quad 0.5$ are quite different, as shown 
in Fig. \ref{fig5}, indicating  the presence of different ``attractors" (other than the 
one corresponding to the stored pattern) with their basins of attraction. Thus we have a complex 
phase-space structure for 
such values of $J_0$. This is consistent with the known result \cite{fischerhertz91} that the
zero-temperature ferromagnetic phase of the SK model is glassy with broken replica symmetry.
On the other hand, for larger values of the acquisition strength, 
e.g. for $J_0=2$, different initial values of $m$ converge to the same $m_{\infty}$, 
indicating a relatively simple structure of the phase space. In Fig. \ref{fig6} we show the evolution of 
$m$ for initial overlaps ranging from $m_0=0.005$ to  $m_0=1$ for $J_0=2$. 
Since the initial overlap $m_0=0.005$ is already very close to the estimated value 
of $\Delta m=0.001$, it appears that all initial configurations with nonzero $m_0$ converge 
to the same attractor that corresponds to the stored pattern. 
Such a large basin of attraction for sufficiently high values of the acquisition strength 
may be an artifact of the 
one pattern model. In the case of many stored patterns, the size of the basin of attraction 
of one of the patterns would get reduced. 
However, the simple one-parameter model brings out the essential feature that it is possible 
to tailor the size of the basin of attraction of a stored pattern by varying the corresponding 
acquisition strength. Note that our result for a simple phase-space structure for large
values of $J_0$ applies only to the subspace of states with finite values of $m$. It does
not preclude the occurrence of a complex phase-space structure in the large subspace of states
with zero magnetization. 

\begin{figure}[htb]
\begin{center}
\epsfxsize=10cm
\epsfysize=8cm
\epsfbox{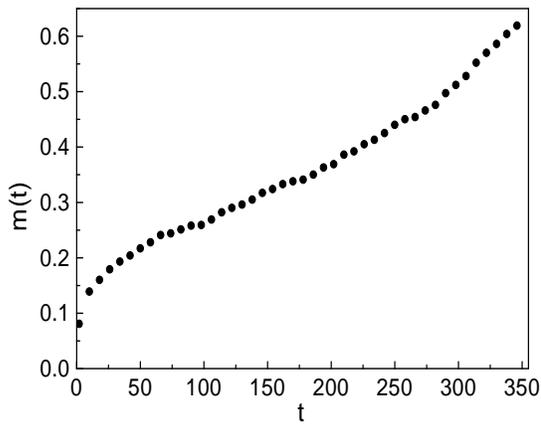}
\end{center}
\caption{\label{fig7} Evolution of the overlap $m(t)$ at even times for $J_0=1.5$, 
$\eta=1$ and $m_0=0.05$.}
\end{figure}

As mentioned earlier, the convergence time, which is the time taken by the network to reach the 
attractor corresponding to the stored pattern from an initial state in the basin of attraction 
of the attractor, is an important parameter in characterizing the performance of the network 
as an associative memory. From studies of spin-glass models, it is known that the 
dynamics for $\eta=1$ may become very slow, especially for small values of the initial overlap 
$m_0$ \cite{ch5_parisibook}. We also find evidence for slow dynamics in our calculations. 
For example, in Fig. \ref{fig7} we show $m(t)$ for $J_0=1.5$ and $m_0=0.05$. 
Even at $t=350$, $m(t)$ does not show any sign of saturation. As there is no 
definite trend in the behavior of $m(t)$, it is not possible to predict the value of $m_{\infty}$ 
and the corresponding time scale.

\subsection{Asymmetric Couplings: $\eta < 1$}
As in the case of $\eta=1$, we first look at the effect of asymmetry in the couplings (by 
lowering the value of $\eta$) on the dynamics with the initial condition $m_0=1$. We find 
that the nature of relaxation changes from a pure power law to a combination of an exponential 
and a power law for $\eta < \eta_{c_1}$. Accordingly, the results of simulations can be 
fitted very well by the function  
\begin{equation}
m(t)=m_{\infty}+\mathrm{const} \times t^{-a}\exp(-t/\tau)\,.
\label{fittingfunction2}
\end{equation}   
Thus for $\eta < \eta_{c_1}$, the overlap decays rapidly to $m_{\infty}$ with a finite 
relaxation time $\tau$. This behavior has also been reported in Ref. \cite{ch5_opper94} for 
$J_0=0$. In that study, the remanent overlap $m_{\infty}$ was found to vanish at the same
value of $\eta$, and the value of $\eta_{c_1}$ was found to be 0.825.  For $J_0 \ne 0$, we find both 
quantitative and qualitative deviations in the behavior of $m(t)$ from those reported in 
Ref. \cite{ch5_opper94}. The value of $\eta_{c_1}$ increases beyond 0.825 as $J_0$ is increased 
from zero. For example for $J_0=1.5$, we have exponential relaxation for values of $\eta$ as high 
as 0.95. This is shown in Fig. \ref{fig9}. Moreover, $m_{\infty}$ vanishes for 
$\eta < \eta_{c_2} < \eta_{c_1}$. It is only at $J_0=0$ that $\eta_{c_2} = \eta_{c_1}$. The 
value of $\eta_{c_2}$ decreases as $J_0$ is increased. For example, for $J_0=0.8$, $\eta_{c_2} \simeq 0.65$ 
whereas for $J_0=1.5$ $\eta_{c_2} \simeq -0.2$. Furthermore, in the ferromagnetic phase, we 
find that over a considerable range of values of $\eta$, there is very little variation in the 
values of $m_{\infty}$, e.g. for $J_0=1.5$, $m_{\infty}$ varies from 0.80 to 0.72 when $\eta$ 
is reduced from 1 to 0. This feature of the network is highly desirable when the possibility 
of functional improvement is explored in the presence of asymmetry in the couplings. It 
ensures that the retrieval quality does not suffer significantly when $\eta < 1$. When $J_0$ is  
sufficiently large, $m_{\infty}$ always remains close to unity, e.g., for $J_0=2$, $m_{\infty}$ 
varies 0.94 to 0.92 when $\eta$ is varied in its full range from 1 to $-1$.

\begin{figure}[htb]
\begin{center}
\epsfxsize=10cm
\epsfysize=8cm
\epsfbox{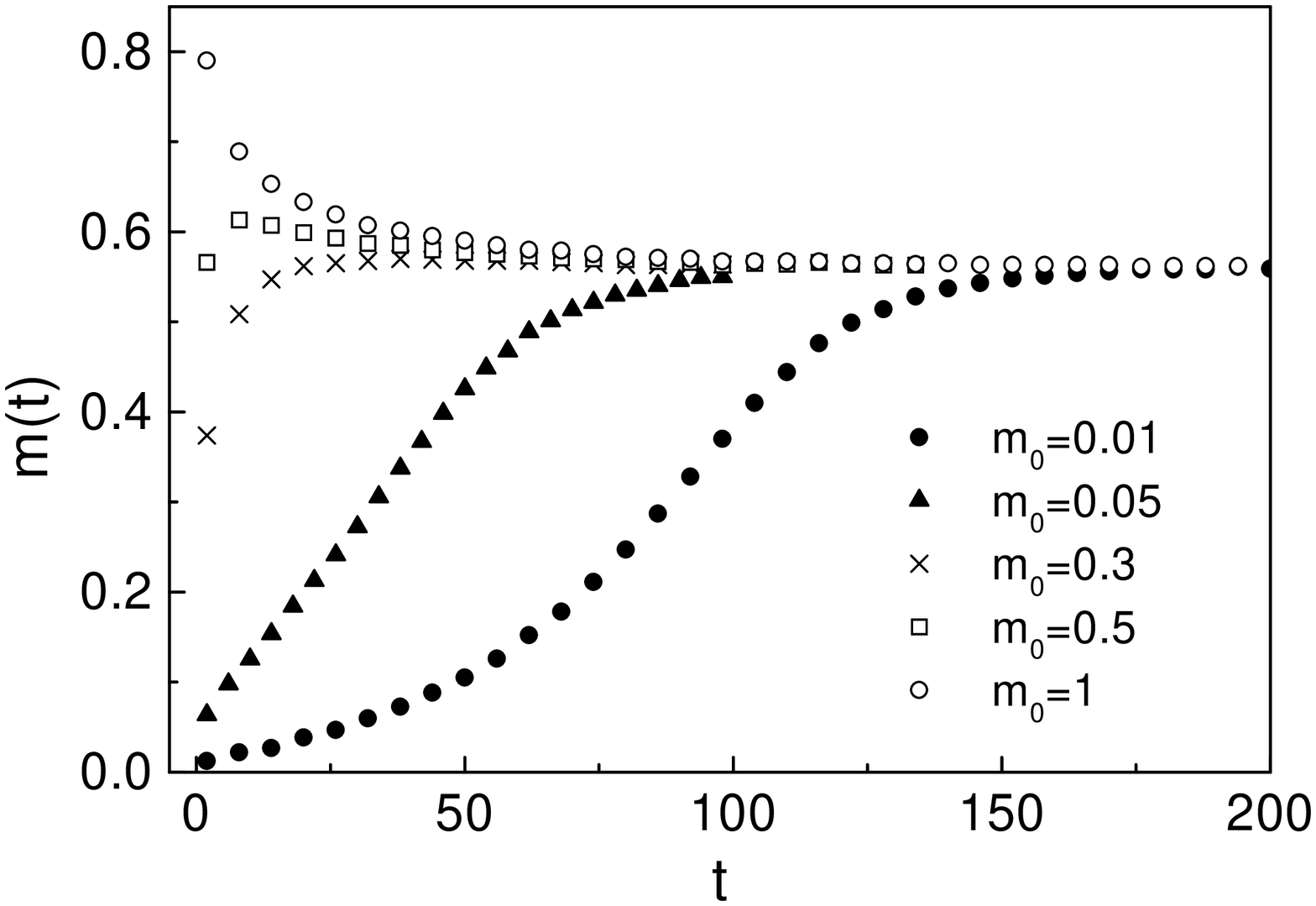}
\end{center}
\caption{\label{fig8} Evolution of the overlap $m(t)$ at even times for different values 
of $m_0$ in a network with $\eta=0.6$ and $J_0=1.3$.}
\end{figure}

\begin{figure}[htb]
\begin{center}
\epsfxsize=10cm
\epsfysize=8cm
\epsfbox{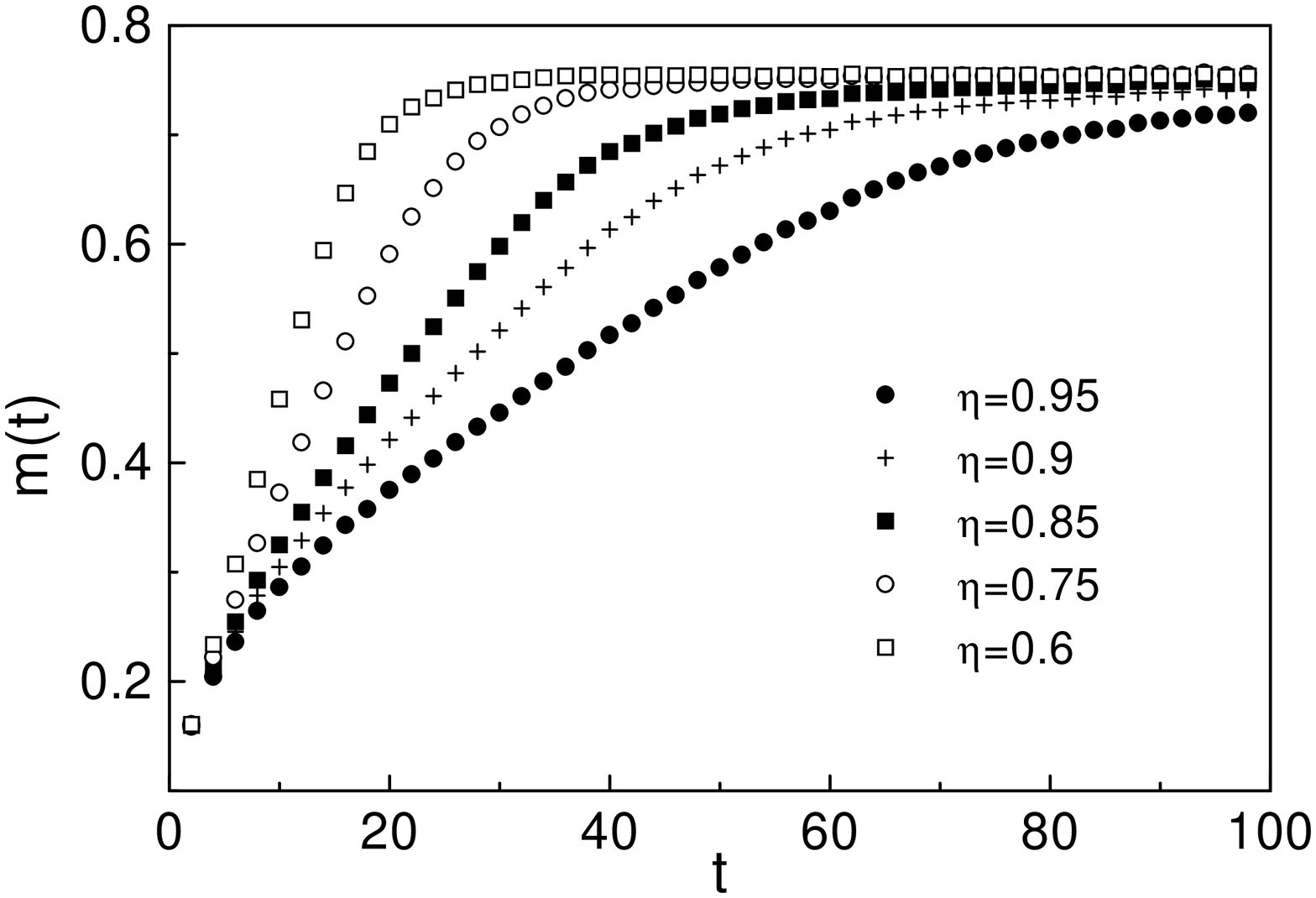}
\end{center}
\caption{\label{fig9} Evolution of the overlap $m(t)$ at even times for different values of 
$\eta$. The initial overlap $m_0$ is 0.1 and $J_0=1.5$.}
\end{figure}

Studies of models of spin glasses and neural networks suggest that the presence of asymmetry 
of an appropriate magnitude in the synaptic interaction matrix may result in the improvement of 
the performance of the network as an associative memory (see Ref. \cite{us1} for a detailed discussion 
on this aspect). It is expected that the asymmetry may destabilize some of the spurious attractors 
which do not correspond any stored pattern. If this happens, then the basin of attraction of 
the stored patterns would increase in size and the retrieval of memory would become faster. We 
find evidence of both of these effects. In Fig. \ref{fig8}, we show the evolution of 
the overlap $m(t)$ for various values of initial overlap ranging from $m_0=0.01$ to $m_0=1$ 
for $J_0=1.3$ and $\eta=0.6$. It can be easily seen that all these initial states converge to 
the same attractor  with  $m_{\infty} \simeq 0.56$. This signifies a considerable enhancement 
in the size of the basin of 
attraction when we compare this behaviour with that in Fig. \ref{fig5}. 
Note, however, that the retrieval quality 
has degraded in the model with synaptic asymmetry: the attractor corresponding the 
stored pattern has 68$\%$ overlap with the pattern for $\eta=1$ compared to the $\simeq 56\%$ 
overlap for $\eta=0.6$. 

In Fig. \ref{fig9}, we plot $m(t)$ for $J_0=1.5$ and $m_0=0.1$ for various values of 
the asymmetry parameter $\eta$. 
It is very clear that the retrieval of memory becomes faster as the asymmetry in couplings 
is increased. By fitting $m(t)$ with the function given in Eq. (\ref{fittingfunction2}), we find 
that the time constant $\tau$ reduces from 55.56 to 4.61 when $\eta$ is varied from 0.95 to 
0.6. At the same time, the value of the final overlap $m_\infty$ does not change much, indicating
that the quality of retrieval is not substantially affected by the introduction of asymmetry in
the synaptic interactions.
   
\section{Conclusion and Discussions}
\label{conclusion}
To summarize, we have studied the synchronous dynamics of a one-pattern model of associative 
memory using a mean field Monte Carlo method. 
Though simple, the model embodies sufficient richness to be useful in predicting the 
behavior of some of the more complicated models of associative memory such as the asymmetric 
Hopfield model. The two relevant parameters in the model are the acquisition strength $J_0$
of the stored pattern and the symmetry parameter $\eta$.
For symmetric couplings ($\eta=1$), we find evidence for a transition at $J_0=1$. 
For $J_0 < 1$, we have a spin glass phase in which the retrieval overlap with the stored pattern 
is small, arising from a ``remanence effect" (non-ergodic relaxation of the system through a 
complex energy landscape). On the other hand, for $J_0 > 1$ we have a ferromagnetic phase 
where the retrieval overlap with the stored pattern increases rapidly with $J_0$ and becomes 
very close to unity. Values of $J_0 \ge 1.3$ would be required for a reasonable 
retrieval of the memory. Inside the ferromagnetic phase, we find the existence of two types of 
phase-space structure in the subspace of states having nonzero overlaps with the stored pattern. 
For example, for $J_0=2$, we have a relatively simple phase-space structure 
where all initial states with a nonzero overlap with the stored pattern flow to the attractor 
corresponding to the stored pattern. In contrast, for $J_0 =1.3$ there are many attractors 
with their own basins of attraction. We also find that the process of retrieval becomes very  
slow for small values of $m_0$. When random asymmetry is introduced in the 
couplings ($\eta < 1$), we find that it results, in general, in better retrieval performance of 
the network by enhancing the size of the basin of attraction of the stored pattern, as well as 
by making the recall of memory significantly faster. 

How do the results described above compare with those obtained for the Hopfield  model with 
random asymmetric interactions \cite{us1,us2}?  Numerical simulations in Ref. \cite{us2} have shown 
that the  presence of asymmetry in the synaptic interaction matrix makes the convergence to spurious 
attractors slower. On the other hand, the convergence time for correct retrieval is only marginally increased 
in the presence of asymmetry. Thus, asymmetry in the synaptic interaction matrix  enhances the performance 
of the Hopfield net as an associative memory by providing a way of discriminating between spurious and 
retrieval attractors by looking at the dynamics of the network. In the model studied here, the improvement in 
performance occurs in a more direct manner: the retrieval becomes faster in the presence 
of asymmetry in the synaptic interaction matrix. Furthermore, the simulation results of 
Ref. \cite{us2} show that the introduction of asymmetry in the synaptic interaction matrix does 
not cause any enhancement of the typical size of the basins of attraction of stored patterns in the 
Hopfield model. Here we do find an enhancement of the size of the basin of attraction of the stored 
pattern. However, this occurs for a very restricted range of values of the parameter $J_0$. These results may 
be indicative of the enlargement of the basin of attraction being a model specific feature. 
Confirmation of some of these issues by extending the method used here to the asymmetric Hopfield model 
and to other models using different learning rules would be most interesting. A somewhat straightforward 
generalization of the technique for the case of two stored patterns 
with different acquisition strengths would provide useful insights into how the storage of 
other patterns, e.g., in the Hopfield model, would modify the results described above. 
Furthermore, this method may also be useful in analysing the effect of asymmetry in the synaptic interaction 
matrix of models of short-term memory \cite{ch5_tdc,mezard86}.

\end{document}